\documentclass[pdftex, twocolappendix, numberedappendix, iop, apj]{emulateapj}

\usepackage{hyperref}
\usepackage{graphicx}
\usepackage{color}
\usepackage{natbib}
\usepackage{sidecap}
\usepackage{enumitem}
\usepackage{amsmath}
\usepackage{amssymb}
\usepackage{courier}

\newcommand{\be}{\begin{equation}}
\newcommand{\ee}{\end{equation}}

\newcommand{\ba}{\begin{eqnarray}}
\newcommand{\ea}{\end{eqnarray}}

\newcommand{\wmap}{\textsl{WMAP}}

\newcommand{\planck}{\textsl{Planck}}
\newcommand{\Planck}{\textsl{Planck}}
\newcommand{\lcdm}{\ensuremath{\Lambda\mathrm{CDM}}}
\newcommand{\pico}{\texttt{PICO}}
\newcommand{\camb}{\texttt{CAMB}}

\slugcomment{\it{}Submitted to the Astrophysical Journal, 30 October 2015}

\shorttitle{2015 \textsl{Planck} Discordance}
\shortauthors{G.~E.~Addison et al.} 

\begin{document}

\title{Quantifying Discordance in the 2015 \textsl{Planck} CMB spectrum}

\author{G.~E.~Addison\altaffilmark{1}, Y.~Huang\altaffilmark{1}, D.~J.~Watts\altaffilmark{1}, C.~L.~Bennett\altaffilmark{1}, M.~Halpern\altaffilmark{2}, G.~Hinshaw\altaffilmark{2},  and J.~L.~Weiland\altaffilmark{1}
}

\email{gaddison@jhu.edu}

\altaffiltext{1}{
Dept. of Physics \& Astronomy, The John Hopkins University, 3400 N. Charles St., Baltimore, MD 21218-2686
}

\altaffiltext{2}{
Department of Physics and Astronomy, University of British Columbia, 6224 Agricultural Road, Vancouver, BC V6T 1Z1, Canada
}

\begin{abstract}

We examine the internal consistency of the \planck\ 2015 cosmic microwave background (CMB) temperature anisotropy power spectrum. We show that tension exists between cosmological constant cold dark matter (\lcdm) model parameters inferred from multipoles $\ell<1000$ (roughly those accessible to \textsl{Wilkinson Microwave Anisotropy Probe}), and from $\ell\geq1000$, particularly the CDM density, $\Omega_ch^2$, which is discrepant at $2.5\sigma$ for a \planck-motivated prior on the optical depth, $\tau=0.07\pm0.02$. We find some parameter tensions to be larger than previously reported because of inaccuracy in the code used by the \planck\ Collaboration to generate model spectra. The \planck\ $\ell\geq1000$ constraints are also in tension with low-redshift data sets, including \planck's own measurement of the CMB lensing power spectrum ($2.4\sigma$), and the most precise baryon acoustic oscillation (BAO) scale determination ($2.5\sigma$). The Hubble constant predicted by \planck\ from $\ell\geq1000$, $H_0=64.1\pm1.7~$km$~$s$^{-1}~$Mpc$^{-1}$, disagrees with the most precise local distance ladder measurement of $73.0\pm2.4~$km$~$s$^{-1}~$Mpc$^{-1}$ at the $3.0\sigma$ level, while the \planck\ value from $\ell<1000$, $69.7\pm1.7~$km$~$s$^{-1}~$Mpc$^{-1}$, is consistent within $1\sigma$. A discrepancy between the \planck\ and South Pole Telescope (SPT) high-multipole CMB spectra disfavors interpreting these tensions as evidence for new physics. We conclude that the parameters from the \planck\ high-multipole spectrum probably differ from the underlying values due to either an unlikely statistical fluctuation or unaccounted-for systematics persisting in the \planck\ data.
\end{abstract}


\section{Introduction}

Measurements of the power spectrum of cosmic microwave background (CMB) temperature fluctuations (hereafter `TT spectrum') are a cornerstone of modern cosmology. The most precise constraints are currently provided by the final 9-year \textsl{Wilkinson Microwave Anisotropy Probe} (\wmap) analysis \citep{bennett/etal:2013,hinshaw/etal:2013}, high-resolution ground-based instruments including the Atacama Cosmology Telescope \citep[ACT;][]{sievers/etal:2013} and the South Pole Telescope \citep[SPT;][]{story/etal:2013}, and most recently \planck\ \citep{planck/13:2015}. Significant improvements in both CMB polarization and low-redshift, late-time observations are anticipated in the near future and will be used to measure or tightly constrain key cosmological quantities including the total neutrino mass, deviations of dark energy from a cosmological constant and the amplitude of primordial gravitational waves \citep[e.g.,][]{snowmass/inflation:2013,snowmass/neutrino:2013,snowmass/darkenergy:2013}. Many of these future results will rely on having precise and accurate TT constraints. Assessing consistency both between and internally within each TT measurement is therefore extremely important. 

While the \planck\ data from the first data release in 2013 \citep{planck/16:2013} were qualitatively in agreement with \wmap, supporting the minimal \lcdm\ model, there were small but highly significant quantitative differences between the cosmological parameters inferred. For example, \citet{larson/etal:2015} found a $\sim6\sigma$ overall parameter discrepancy after accounting for the cosmic variance common to both experiments.

Several systematic effects were corrected in the \planck\ 2015 data release, including issues relating to data calibration and map making \citep{planck/01:2015}, which led to a shift in the inferred TT power spectrum amplitude by $3.5\sigma$ in units of the 2015 uncertainty \citep[Table~1 of][]{planck/13:2015}, and an artifact with a statistical significance of $2.4-3.1\sigma$ near multipole $\ell\simeq1800$ in the 217~GHz temperature power spectrum \citep{planck/12:2013}. See also discussion in \citet{spergel/etal:2015}.

The \wmap\ and \planck\ 2015 TT spectra now appear to be in agreement over their common multipole range \citep[Fig.~46 of][]{planck/11:2015}. When the additional information in the high-order acoustic peaks and damping tail of the TT spectrum are included, however, the \planck\ parameters pull away from \wmap\ \citep[Section 4.1.6 of][]{planck/11:2015}, leading to tension between \planck\ and several low-redshift cosmological measurements if \lcdm\ is assumed, including a $2.5\sigma$ tension with the \citet{riess/etal:2011} determination of the Hubble constant, $H_0$, $2-3\sigma$ tension with weak lensing measurements of the CFHTLens survey \citep{heymans/etal:2012}, and tension with the abundance of massive galaxy clusters \citep[e.g.,][]{planck/24:2015}.

In this paper we examine the internal consistency of the \planck\ TT spectrum. We show that tension exists between $\lcdm$ parameters inferred from the \planck\ TT spectrum at the multipoles accessible to \wmap\ ($\ell\lesssim1000$) and at higher multipoles ($\ell\gtrsim1000$). The constraints from high multipoles are, furthermore, in tension with many low-redshift cosmological measurements, including \planck's own lensing potential power spectrum measurement and baryon acoustic oscillation (BAO) from galaxy surveys, while the low-multipole \planck\ TT, \planck\ lensing, \wmap, BAO, and distance ladder $H_0$ data are all in reasonable agreement.

We describe the data sets used and parameter fitting methodology in Section~2 and present results in Section~3. Discussions and conclusions follow in Sections~4 and 5.

\section{Data and parameter fitting}

We use \camb\footnote{\url{camb.info}} \citep{lewis/etal:2000} to calculate temperature and lensing potential power spectra as a function of cosmological parameters and \texttt{CosmoMC}\footnote{\url{http://cosmologist.info/cosmomc/}} \citep{lewis/bridle:2002} to perform Monte-Carlo Markov Chain (MCMC) parameter fitting and obtain marginalized parameter distributions, adopting the default \planck\ settings, including a neutrino mass of 0.06~eV \citep{planck/16:2013}. We use the public temperature-only \planck\ 2015 \texttt{lowl} likelihood for $2\leq\ell\leq29$, the binned \texttt{plik} likelihood for $30\leq\ell\leq2508$, and, in some cases, the \planck\ 2015 lensing likelihood, which includes multipoles of the lensing potential power spectrum $C_L^{\phi\phi}$ covering $40\leq L\leq400$ \citep{planck/11:2015,planck/15:2015}. We fit for six \lcdm\ parameters: the physical baryon and CDM densities, $\Omega_bh^2$ and $\Omega_ch^2$, the angular acoustic scale, parametrized by $\theta_{\rm MC}$, the optical depth, $\tau$, the primordial scalar fluctuation amplitude, $A_s$, and the scalar spectral index, $n_s$. Other parameters, including $H_0$, the total matter density, $\Omega_m$, and the present-day mass fluctuation amplitude, $\sigma_8$, are derived from these six. Additional foreground and calibration parameters used in the fits are described by \citet{planck/11:2015}.

At the present time the analysis of \planck's polarization data is only partially complete. At high multipoles, significant systematic errors remain in the TE and EE spectra, putatively due to beam mismatch, which leads to temperature-polarization leakage \citep[Sec.~3.3.2 of][]{planck/13:2015}.  At low multipoles ($\ell<30$), the 100, 143 and 217~GHz polarization data have significant residual systematic errors and are ``not considered usable for cosmological analyses''\footnote{According to the \Planck\ 2015 Release Explanatory Supplement \url{http://wiki.cosmos.esa.int/planckpla2015/index.php/Frequency_Maps\#Caveats_and_known_issues}}. The LFI 70~GHz data, in conjunction with the 30 and 353~GHz maps as Galactic foreground tracers, are used to constrain $\tau$. Using the polarized 353~GHz map as a dust tracer results in a value of $\tau$ lower than constraints from \wmap\ \citep[$0.066\pm0.016$ compared to $0.089\pm0.014$,][]{hinshaw/etal:2013,planck/13:2015}. Given these complexities and uncertainties, we have chosen to leave polarization data out of the current analysis and focus on conclusions that can be drawn from the TT data alone.

Without polarization data, $\tau$ is only weakly constrained, but it does couple to other cosmological parameters. We considered two approaches for setting priors on $\tau$. First we adopted a Gaussian prior of $\tau=0.07\pm0.02$ as in \citet{planck/11:2015}, which is consistent within $1\sigma$ with the range of values inferred from \wmap\ and \planck\ data \citep{hinshaw/etal:2013,planck/13:2015}. Second, to gain more insight into exactly how $\tau$ does or does not affect our conclusions about TT consistency, we also ran chains with $\tau$ fixed to specific values: 0.06, 0.07, 0.08, and 0.09.

When assessing consistency between parameter constraints from two data sets that can be considered independent we use the difference of mean parameter values, which we treat as multivariate Gaussian with zero mean and covariance given by the sum of the covariance matrices from the individual data sets. The mean and covariance for each data set are estimated from the MCMC chains. We then quote equivalent Gaussian `sigma' levels for the significance of the parameter differences.

We also considered using the difference of best-fit parameters, rather than difference of means, for these comparisons. For Gaussian posterior distributions this choice should make little difference. We find that this is generally true, with significance levels for parameter differences changing only at the $0.1-0.2\sigma$ level. In a few cases, however, we found a significant shift, due to an offset between the mean and best-fit parameters. In all cases the Gaussian distribution specified by the mean and covariance matrix from the chains provided an excellent match to the distribution of the actual MCMC samples, and for this reason we quote results based on the differences of the mean rather than best-fit parameters. It is possible that the mismatches are caused by problems in the algorithm used to determine the best-fit parameters\footnote{See \url{http://cosmologist.info/cosmomc/readme.html}}. Note that simply taking the maximum-likelihood parameters directly from the MCMC chains is unreliable due to the large parameter volume sampled (typically around 20 parameters, including nuisance parameters, e.g., for foregrounds). The overall posterior distribution is well mapped out by a converged chain but the tiny region of parameter space close to the likelihood peak is not.

\section{Results}

Figure~1 shows the two-dimensional \lcdm\ parameter constraints for the \planck\ 2015 TT spectra spanning $2\leq\ell<1000$ and $1000\leq\ell\leq2508$, with a $\tau=0.07\pm0.02$ prior. Similar contours are shown in Figure~31 of \citet{planck/11:2015} using the same prior on $\tau$. Two differences in our fit act to pull some of the low and high multipole parameter constraints away from one another. Firstly, the constraints in the \planck\ figure only extend down to $\ell=30$ because the intention was to test robustness of the \texttt{plik} likelihood only.  We use the full range $2\leq\ell<1000$ with the intention of examining parameter values. Secondly, the \planck\ fit uses the \pico\footnote{\url{https://pypi.python.org/pypi/pypico}} \citep{fendt/wandelt:2007a} code rather than \camb\ to generate TT spectra. We find that the \pico\ and \camb\ results are noticeably different for the $1000\leq\ell\leq2508$ fit. \pico\ requires only a fraction of the computation time and provides a good approximation to \camb, but only within a limited volume of parameter space. Some parameter combinations outside this volume are allowed by the $1000\leq\ell\leq2508$ data. In these cases, the \pico\ output deviates from the \camb\ spectrum and a poor likelihood is returned, leading to artificial truncation of the contours, particularly for $\Omega_bh^2$ and $n_s$.

\begin{figure*}
	\centering
	\includegraphics[width=180mm]{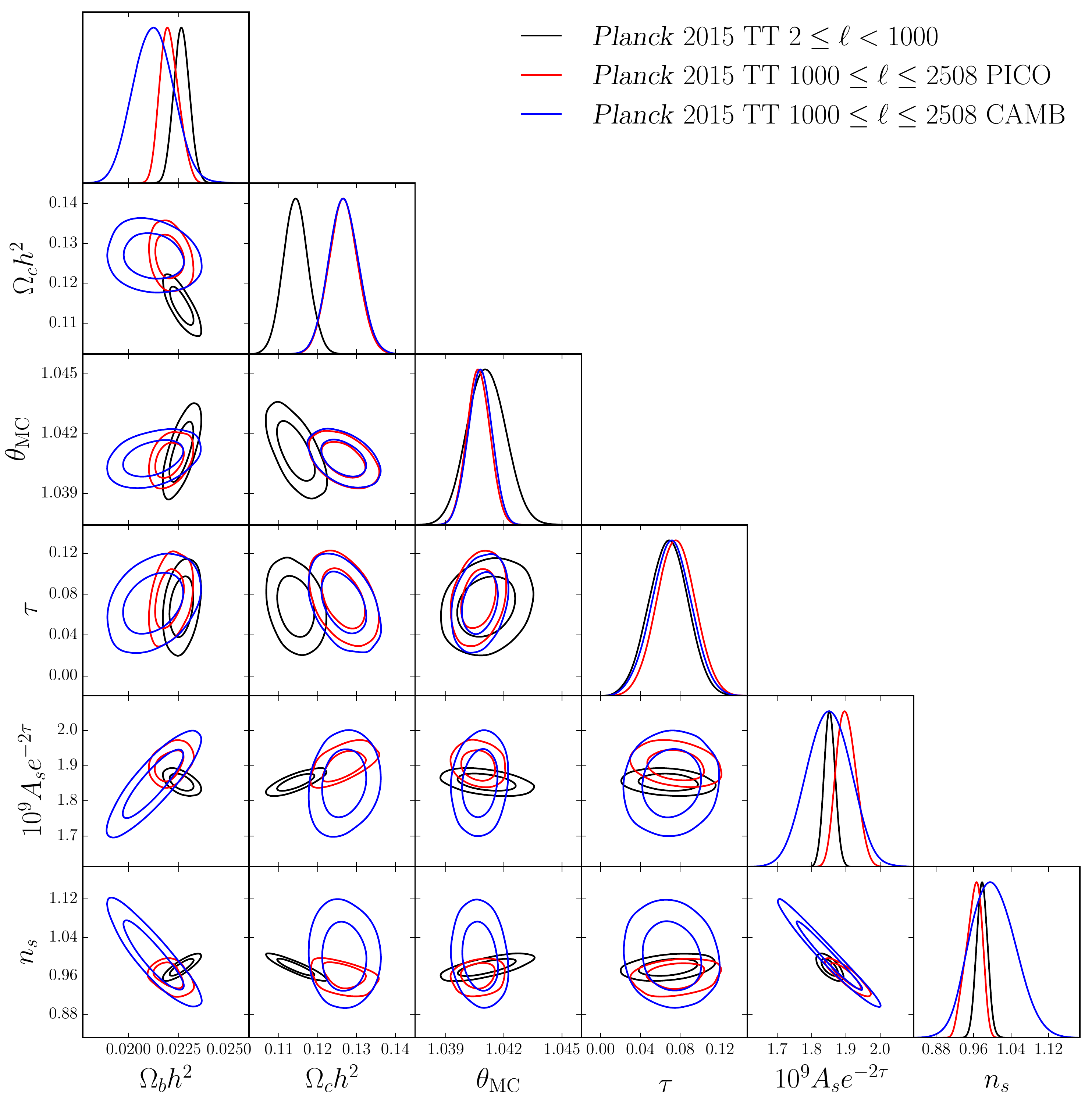}
	\caption{Contours enclosing 68.3\% and 95.5\% of MCMC sample points from fits to the \planck\ TT spectrum. Results are shown for $2\leq\ell<1000$, roughly the multipole range accessible to \wmap, and higher multipoles, $1000\leq\ell\leq2508$. These constraints are effectively independent and are in tension, for example $\Omega_ch^2$ differs by $2.5\sigma$. Results are also shown for the $1000\leq\ell\leq2508$ fit where the \pico\ code is used to estimate the theoretical TT spectra instead of the more accurate \camb. Using \pico\ leads to an artificial truncation of the contours and diminishes the discrepancy between the high and low multipole fits for some parameters. We adopt a Gaussian prior of $\tau=0.07\pm0.02$.}
\end{figure*}

From Figure~1 it is clear that some tension exists between parameters inferred from the $\ell<1000$ and $\ell\geq1000$ \planck\ TT spectra. Assuming the two sets of constraints are independent, the values of $\Omega_ch^2$ differ by $2.5\sigma$. Independence is a valid assumption because even the bins on either side of the $\ell=1000$ split point are only correlated at the 4\% level and the degree of correlation falls off with increasing bin separation. Taken together the five free $\lcdm$ parameters differ by $1.8\sigma$, however it should be noted that $\Omega_ch^2$ plays a far more significant role in comparisons with low-redshift cosmological constraints (Section 3.3) than, for example, $\theta_{\rm MC}$.

For fixed $\tau$ we find differences in $\Omega_ch^2$ of $3.0$, $2.7$, $2.9$, and $2.1\sigma$ for $\tau$ values of 0.06, 0.07, 0.08 and 0.09, respectively. Constraints on each parameter for these cases are shown in Figure~2. Apart from the expected strong correlation with $A_s$ (the TT power spectrum amplitude scales as $A_se^{-2\tau}$) there is relatively little variation with $\tau$. Note that while increasing $\tau$ reduces the tension in $\Omega_ch^2$, higher values of $\tau$ are mildly disfavored by \planck's own polarization analysis \citep{planck/13:2015}.

\begin{figure*}
	\centering
	\includegraphics[width=180mm]{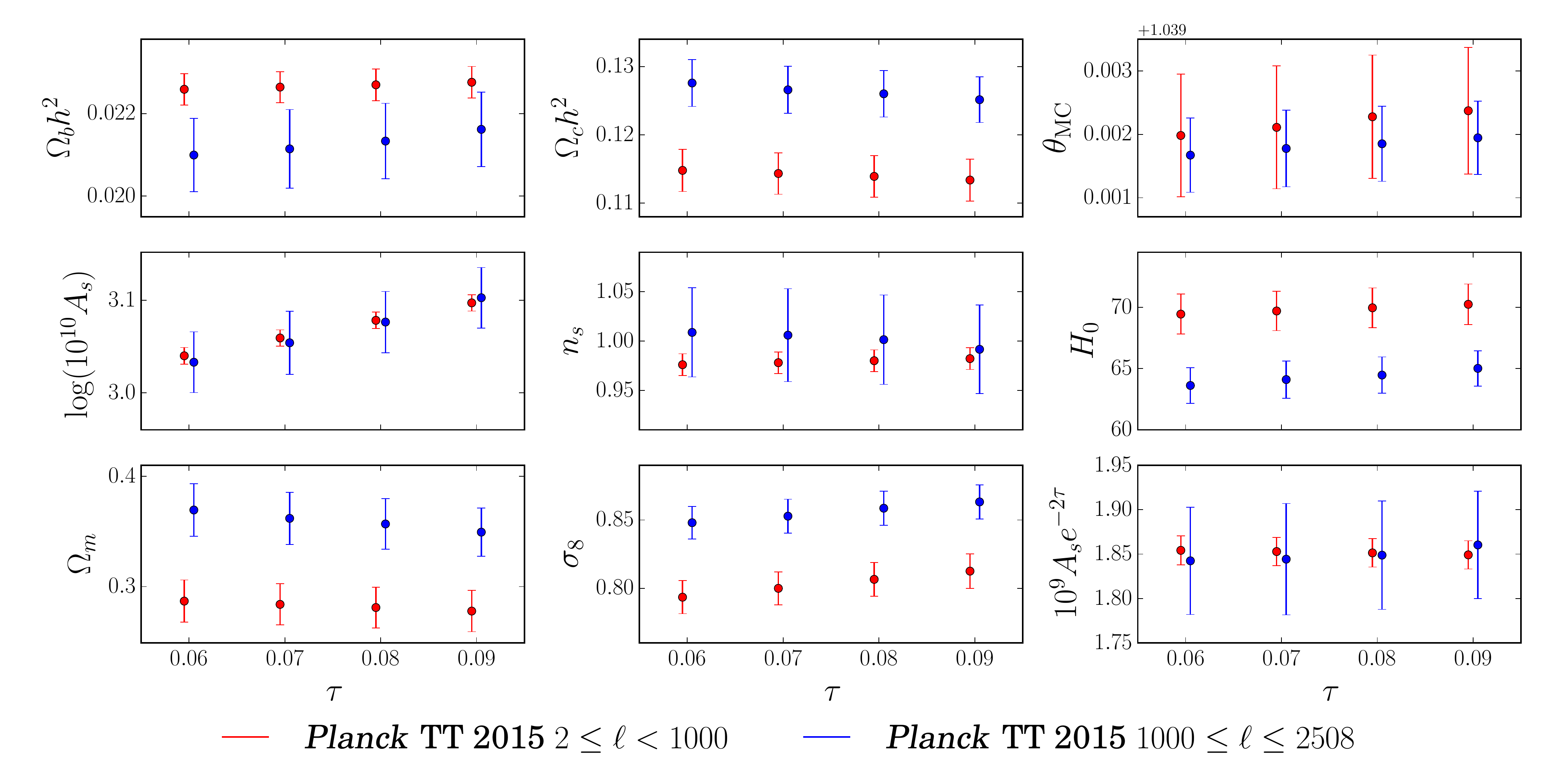}
	\caption{Marginalized 68.3\% confidence \lcdm\ parameter constraints from fits to the $\ell<1000$ and $\ell\geq1000$ \planck\ TT spectra. Here we replace the prior on $\tau$ with fixed values of 0.06, 0.07, 0.08, and 0.09, to more clearly assess the effect $\tau$ has on other parameters in these fits. Aside from the strong correlation with $A_s$, which arises because the TT spectrum amplitude scales as $A_se^{-2\tau}$, dependence on $\tau$ is fairly weak. Tension at the $>2\sigma$ level is apparent in $\Omega_ch^2$ and derived parameters, including $H_0$, $\Omega_m$, and $\sigma_8$.}
\end{figure*}

We investigated the effect of fixing the foreground parameters to the best-fit values inferred from the fit to the whole \planck\ multipole range rather than allowing them to vary separately in the $\ell<1000$ and $\ell\geq1000$ fits. This helps break degeneracies between foreground and \lcdm\ parameters and leads to small shifts in \lcdm\ parameter agreement, with the tension in $\Omega_ch^2$ decreasing to $2.3\sigma$ for $\tau=0.07\pm0.02$, for example. The best-fit $\chi^2$ is, however, worse by 3.1 and 4.8 for the $\ell<1000$ and $\ell\geq1000$ fits, respectively, reflecting the fact that the $\ell<1000$ and $\ell\geq1000$ data mildly prefer different foreground parameters. Overall the choice of foreground parameters does not significantly impact our conclusions.

\subsection{Comparing temperature and lensing spectra}

\citet{planck/13:2015} found that allowing a non-physical enhancement of the lensing effect in the TT power spectrum, parametrized by the amplitude parameter $A_L$ \citep{calabrese/etal:2008}, was effective at relieving the tension between the low and high multipole \planck\ TT constraints. For the range of scales covered by \planck, the main effect of increasing $A_L$ is to slightly smooth out the acoustic peaks. If \lcdm\ parameters are fixed, a 20\% change in $A_L$ suppresses the fourth and higher peaks by around 0.5\% and raises troughs by around 1\%, for example.

In Figure~3 we show the effect of fixing $A_L$ to values other than the physical value of unity on the $\ell<1000$ and $\ell\geq1000$ parameter comparison, for $\tau=0.07\pm0.02$. For $A_L>1$ the parameters from $\ell\geq1000$ shift toward the $\ell<1000$ results, resulting in lower values of $\Omega_ch^2$ and higher values of $H_0$. \citet{planck/13:2015} found $A_L=1.22\pm0.10$ for \texttt{plik} combined with the low-$\ell$ \planck\ joint temperature and polarization likelihood, although note that this fit was performed using \pico\ rather than \camb, which uses a somewhat different $A_L$ definition.

\begin{figure*}
	\centering
	\includegraphics[width=180mm]{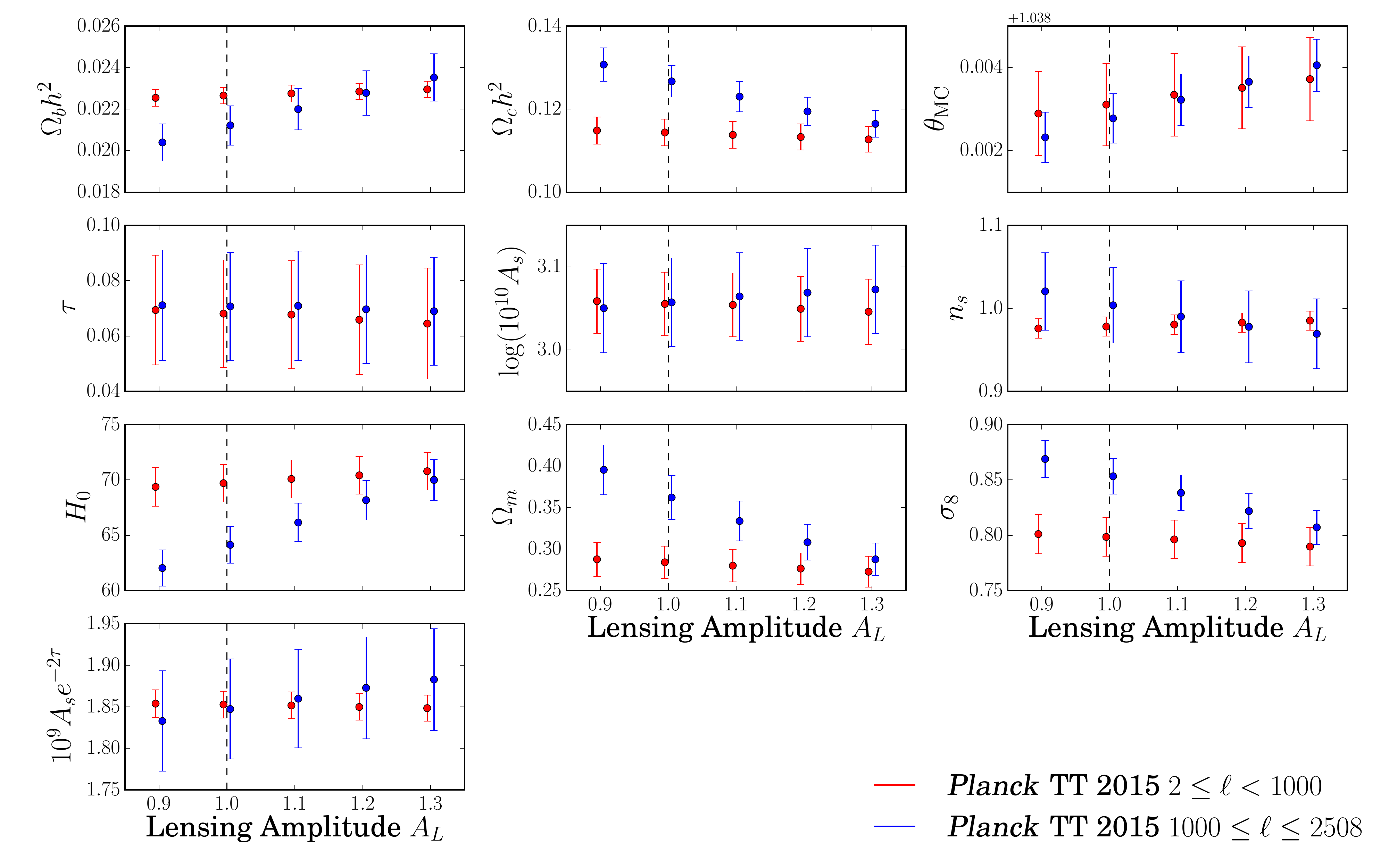}
	\caption{Marginalized 68.3\% parameter constraints from fits to the $\ell<1000$ and $\ell\geq1000$ \planck\ TT spectra with different values of the phenomenological lensing amplitude parameter, $A_L$, which has a physical value of unity (dashed line). Increasing $A_L$ smooths out the high order acoustic peaks, which improves agreement between the two multipole ranges. Note that a high value of $A_L$ is not favored by the direct measurement of the $\phi\phi$ lensing potential power spectrum (see text).}
\end{figure*}

Lensing also induces specific non-Gaussian signatures in CMB maps that can be used to recover the lensing potential power spectrum (hereafter `$\phi\phi$ spectrum'). \citet{planck/15:2015} report a measurement of the $\phi\phi$ spectrum using temperature and polarization data with a combined significance of $\sim40\sigma$. The $\phi\phi$ spectrum constrains $\sigma_8\Omega_m^{0.25}=0.591\pm0.021$, assuming priors of $\Omega_bh^2=0.0223\pm0.0009$, $n_s=0.96\pm0.02$, and $0.4<H_0/100$~km~s$^{-1}$~Mpc$^{-1}<1.0$ \citep{planck/15:2015}. We computed constraints on this same parameter combination from \planck\ TT data using a $\tau=0.07\pm0.02$ prior:
\be
\begin{split}
\sigma_8\Omega_m^{0.25}&=0.591\pm0.021\textrm{ (\planck\ 2015 $\phi\phi$),}\\
&=0.583\pm0.019\textrm{ (\planck\ 2015 TT $\ell<1000$),}\\
&=0.662\pm0.020\textrm{ (\planck\ 2015 TT $\ell\geq1000$)}.
\end{split}
\ee
The $\ell<1000$ and $\ell\geq1000$ TT values differ by $2.9\sigma$, consistent with the difference in $\Omega_ch^2$ discussed above. The $\ell\geq1000$ and $\phi\phi$ values are in tension at the $2.4\sigma$ level (for fixed values of $\tau$ in the range $0.06-0.09$ we find a $2.4-2.5\sigma$ difference). The $\ell<1000$ TT and $\phi\phi$ values are consistent within $0.3\sigma$.

It is worth noting that while allowing $A_L > 1$ does relieve tension between the low-$\ell$ and high-$\ell$ TT results, it does not alleviate the high-$\ell$ TT tension with $\phi\phi$. For $A_L = 1.2$ (by the CAMB definition) we find $\sigma_8\Omega_m^{0.25} = 0.612\pm0.019$ from $\ell\leq1000$, while the $\phi\phi$ spectrum requires $\sigma_8\Omega_m^{0.25} = 0.541\pm0.019$. This is because the $\phi\phi$ power roughly scales as $A_L(\sigma_8\Omega_m^{0.25})^2$, so, for fixed $\phi\phi$, increasing $A_L$ by 20\% requires a $\sim10\%$ decrease in $\sigma_8\Omega_m^{0.25}$. As shown in Figure~4, there is no value of $A_L$ that produces agreement between these data.

\begin{figure}
	\centering
	\includegraphics[width=80mm]{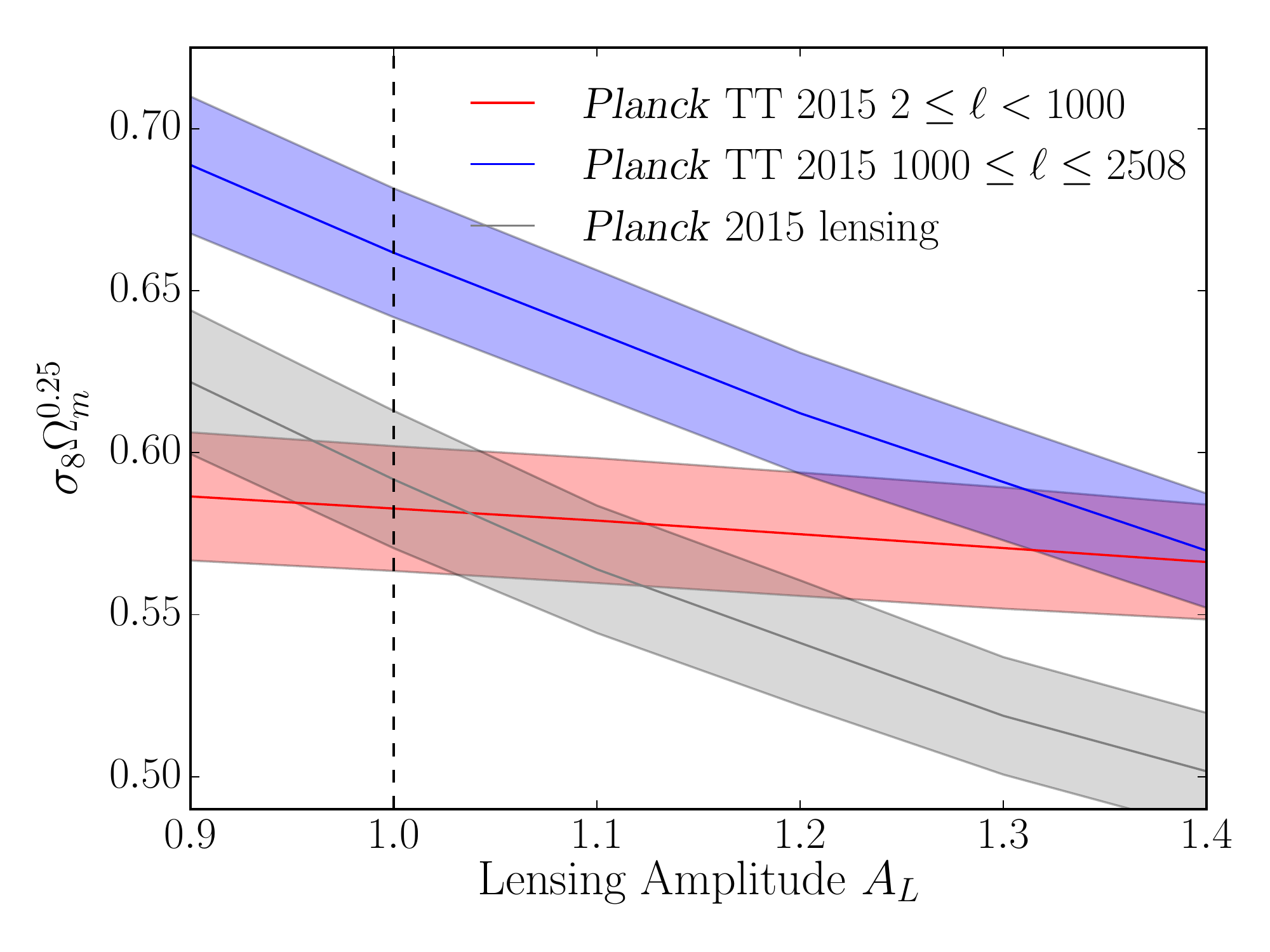}
	\caption{Constraints on $\sigma_8\Omega_m^{0.25}$ from fits to the $\ell<1000$ and $\ell\geq1000$ Planck TT spectra, and to the Planck $\phi\phi$ lensing spectrum.  Results are shown as a function of the phenomenological lensing amplitude parameter $A_L$. The $\phi\phi$ measurement constrains the product $A_L(\sigma_8\Omega_m^{0.25})^2$. A similar trend is apparent in the $\ell\geq1000$ constraint, where lensing has a significant effect. For $\ell<1000$ the lensing effect is small, resulting in almost no dependence on $A_L$. The $\ell<1000$ and $\phi\phi$ constraints agree well for the physical value of $A_L=1$ (dashed line). Increasing $A_L$ helps reconcile the low-$\ell$ and high-$\ell$ constraints but does not improve agreement between the high-$\ell$ and $\phi\phi$ constraints.}
\end{figure}

The $\phi\phi$ spectrum featured prominently in the \planck\ claim that the true value of $\tau$ is lower than the value inferred by \wmap\ \citep{planck/13:2015}. While a full investigation into $\tau$ is deferred to future work we note here that the effect of the $\phi\phi$ spectrum on $\tau$ is completely dependent on the choice of temperature and polarization data. The shift to lower $\tau$ in the joint \planck\ 2015 TT-$\phi\phi$ fit is partly a reflection of the tension discussed above. Adding the \planck\ $\phi\phi$ spectrum to the \wmap9 data, in contrast, leads to no measurable shift in $\tau$ at all, reflecting the fact that the $\phi\phi$ spectrum and \wmap\ temperature and polarization data (with $\tau=0.089\pm0.014$) are in excellent agreement. Figure~5 shows that, while some parameter constraints are tightened by a factor of two over \wmap\ alone, the mean values shift by $<0.25\sigma$.

\begin{figure}
	\centering
	\includegraphics[width=80mm]{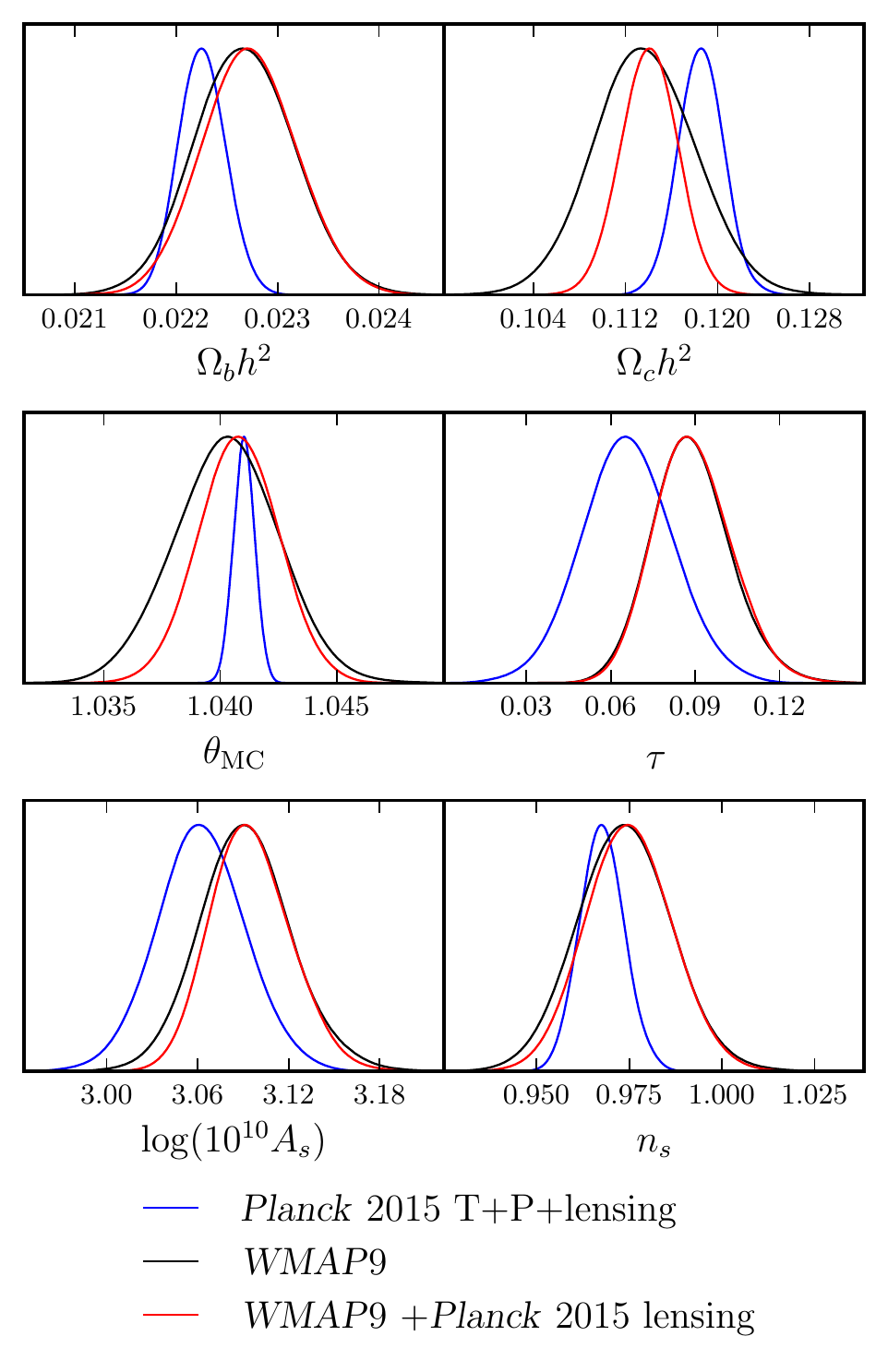}
	\caption{Marginalized \lcdm\ parameter constraints comparing results from \planck\ 2015 (combined temperature, polarization and lensing) with \wmap9 alone and \wmap9 in conjunction with the \planck\ $\phi\phi$ lensing power spectrum. Adding the $\phi\phi$ spectrum to \planck\ temperature and polarization data results in a downward shift in $\tau$, which reflects internal tension between the high-multipole \planck\ TT spectrum and $\phi\phi$ (see text). The \wmap9 and \planck\ $\phi\phi$ constraints are in very good agreement. Adding $\phi\phi$ to \wmap\ leads to a negligible shift in $\tau$ and shifts of $<0.25\sigma$ in other parameters.}
\end{figure}

\subsection{Comparison With SPT}

 \citet{planck/16:2013} reported moderate to strong tension between cosmological parameters from the SPT TT spectrum, measured over 2500 square degrees and covering $650\leq\ell\leq3000$ \citep{story/etal:2013}, and the \planck\ TT spectrum. \citet{planck/13:2015} comment that this tension has worsened for the \planck\ 2015 data. A detailed comparison of these data sets is beyond the scope of this work, however we note that when we recalibrate the public SPT spectrum to the full-sky \planck\ 2015 spectrum following the method described by \citet{story/etal:2013}, using data from $650\leq\ell\leq1000$ and correcting for foregrounds, we recover the original SPT calibration to \wmap\ within $0.3\sigma$. For the 143~GHz \planck\ spectrum, most directly comparable to the 150~GHz SPT channel, the agreement is better than $0.1\sigma$. The disagreement between SPT and \planck\ therefore cannot be resolved by simply calibrating SPT to \planck\ rather than \wmap\ in this manner. We note that the high-multipole ACT TT measurements are consistent with \wmap\ and SPT, as well as \planck\ 2013 if a recalibration is allowed \citep{calabrese/etal:2013,louis/etal:2014}, and so do not currently help our understanding of these tensions. More precise upcoming measurements from ACTPol will be useful for future comparisons.

\subsection{Comparison With BAO and Local $H_0$ Measurements}

Figure~6 shows a comparison of CMB \lcdm\ constraints with the 1\% BAO scale measurement from the Baryon Oscillation Spectroscopic Survey (BOSS) `CMASS' galaxy sample at an effective $z=0.57$ \citep{anderson/etal:2014} and the most precise local distance ladder constraint on the Hubble constant, $H_0=73.0\pm2.4~$km$~$s$^{-1}~$Mpc$^{-1}$ \citep{riess/etal:2011,bennett/etal:2014}. The BAO scale is parametrized as the ratio of the combined radial and transverse dilation scale, $D_V$ \citep{eisenstein/etal:2005}, to the sound horizon at the drag epoch, $r_d$, which has a fiducial value $r_{d\textrm{,fid}}=149.28~$Mpc \citep{anderson/etal:2014}.

The BOSS BAO $D_V/r_d$ constraint is at the higher end of the range preferred by \wmap\ and \planck\ $\ell<1000$, though consistent within $1\sigma$. The \planck\ $\ell\geq1000$ data predict higher values of $D_V/r_d$, and lower values of $H_0$, than the BOSS BAO and distance ladder measurements at the $2.5\sigma$ and $3.0\sigma$ level, respectively, for $\tau=0.07\pm0.02$. The difference between the \planck\ high-multipole constraint and the Riess \emph{et al.} $H_0$ constraint is extremely unlikely to be explained by statistical fluctuation alone. The SPT-only values provided by \citet{story/etal:2013}\footnote{\url{http://pole.uchicago.edu/public/data/story12/chains/}} are also shown. The SPT predictions for $D_V/r_d$ and $H_0$ are discrepant with those from \planck\ $\ell\geq1000$ at the $2.6\sigma$ and $2.7\sigma$ levels. Note that SPT used a \wmap-based $\tau$ prior but that $\tau$ couples very weakly to the inferred BAO scale.

The consistency between the \planck\ and BAO constraints has been repeatedly highlighted \citep{planck/16:2013,planck/13:2015}. We find that this agreement arises more in spite of than because of the high-multipole TT spectrum that \wmap\ did not measure. Figure~7 shows constraints in the $\Omega_m-H_0$ plane for the BAO constraint from combining BOSS CMASS with the BOSS `LOWZ' sample \citep{anderson/etal:2014}, Sloan Digital Sky Survey Main Galaxy Sample \citep[SDSS MGS]{ross/etal:2015}, and Six-degree-Field Galaxy Survey \citep[6dFGS]{beutler/etal:2011} measurements. This is the same combination utilized in the \planck\ 2015 cosmological analysis. The BAO contours are plotted assuming $\Omega_bh^2=0.02223$, although the exact choice has little effect \citep{addison/etal:2013,bennett/etal:2014}. CMB constraints are plotted for comparison. The $>2\sigma$ tension between the \planck\ $\ell\geq1000$ and BOSS BAO constraints persists with the full BAO dataset.

\citet{bennett/etal:2014} combined \wmap9, ACT, SPT, BAO, and distance ladder measurements and found that these measurements are consistent and together constrain $H_0=69.6\pm0.7~$km$~$s$^{-1}~$Mpc$^{-1}$. This concordance value differs from the \planck\ $\ell\geq1000$ constraint of $64.1\pm1.7~$km$~$s$^{-1}~$Mpc$^{-1}$ at $3.1\sigma$ but agrees well with the \planck\ $\ell<1000$ constraint of $69.7\pm1.7~$km$~$s$^{-1}~$Mpc$^{-1}$.

\begin{figure}
	\centering
	\includegraphics[width=80mm]{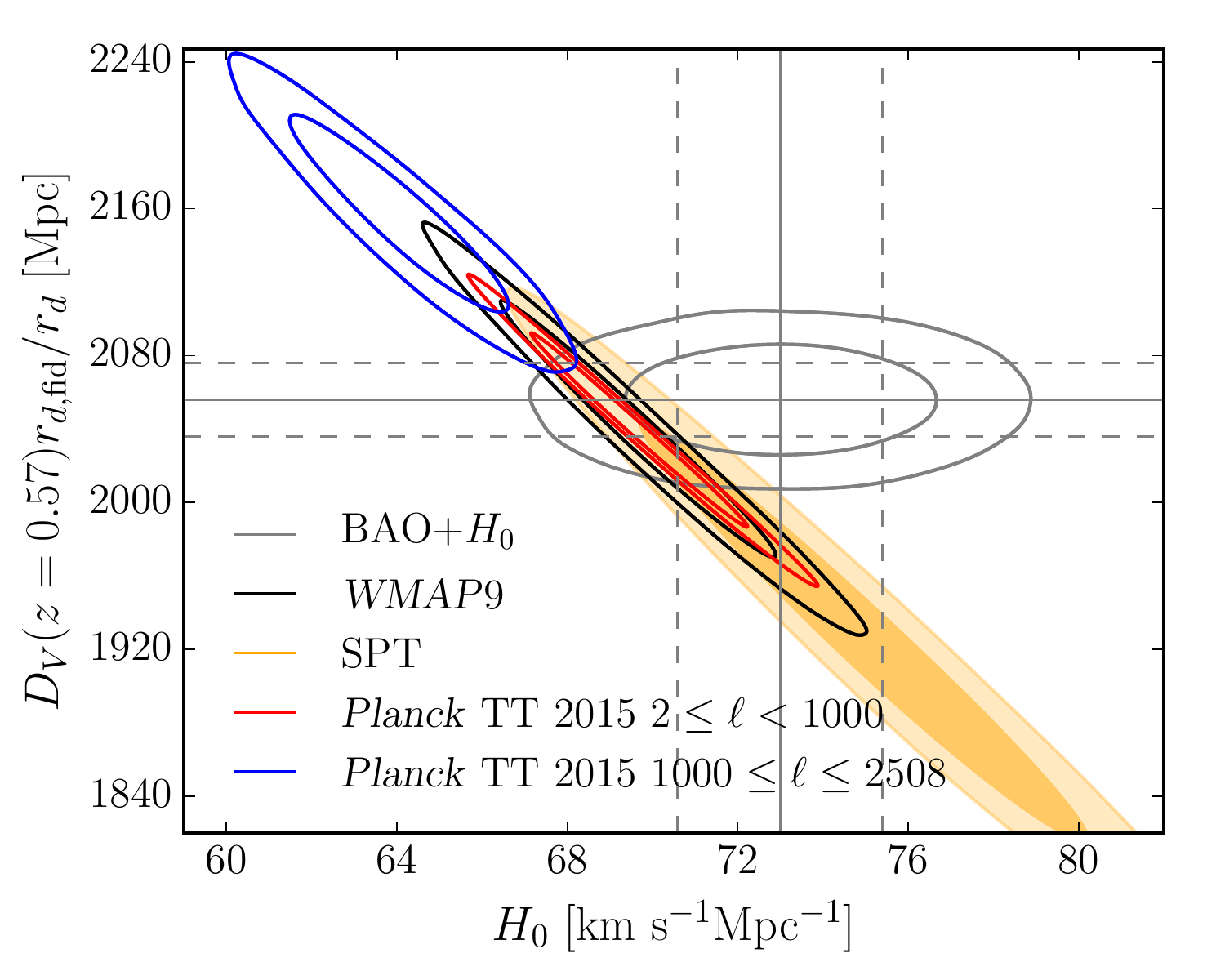}
	\caption{BOSS BAO scale and local distance ladder $H_0$ measurements \citep{riess/etal:2011,anderson/etal:2014,bennett/etal:2014} with \lcdm\ CMB 68.3 and 94.5\% confidence contours overplotted. The \planck\ $\ell\geq1000$ constraints are discrepant with the BAO and distance ladder measurements at the $2.5\sigma$ and $3.0\sigma$ levels, respectively, while the \wmap9 and \planck\ $\ell<1000$ constraints are consistent with both within $1\sigma$. Constraints from SPT (covering $650\leq\ell\leq3000$) are also shown. \Planck\ and SPT currently provide the most precise measurements of the CMB damping tail and their predictions for the $z=0.57$ BAO scale and $H_0$ differ at the $2.6\sigma$ and $2.7\sigma$ level.}
\end{figure}

\begin{figure}
	\centering
	\includegraphics[width=80mm]{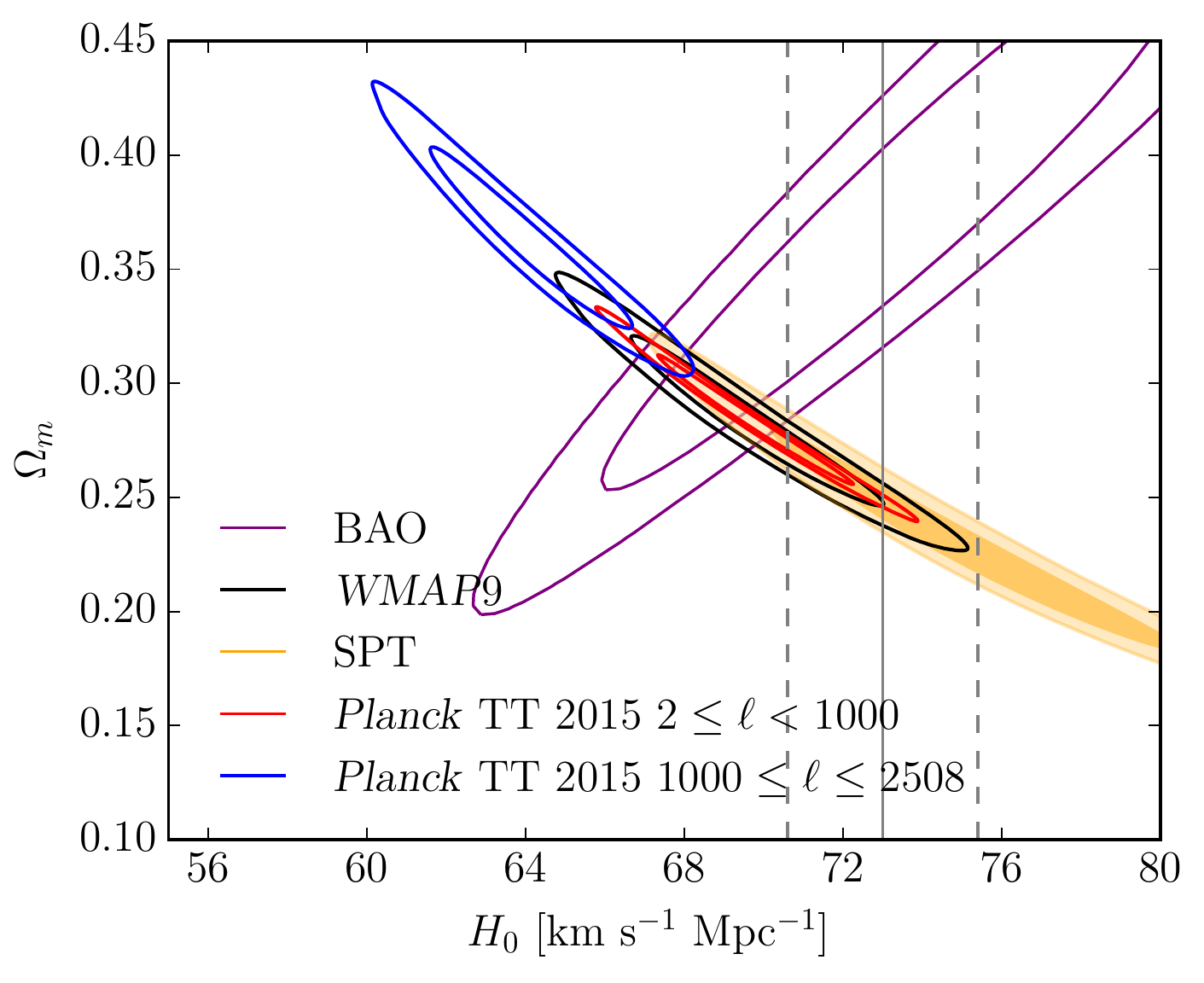}
	\caption{Comparison of CMB, BAO, and distance ladder constraints in the $\Omega_m-H_0$ plane. We show here the BAO constraints from combining the BOSS CMASS, BOSS LOWZ, SDSS MGS, and 6dFGRS measurements, assuming $\Omega_bh^2=0.0223$ (see text). The tension between \planck\ $\ell\geq1000$ and BOSS CMASS BAO (Fig.~6) persists when comparing to the joint BAO constraint.}
\end{figure}

\subsection{Choice of multipole split}

The choice of $\ell=1000$ as the split point for parameter comparisons matches the tests described by \citet{planck/11:2015} and roughly corresponds to the maximum multipoles accessible to \wmap, but the exact choice is arbitrary. To test the robustness of our findings we also considered the effect of splitting the \planck\ TT spectrum at $\ell=800$. This choice achieves an almost-even division of the \planck\ TT spectrum constraining power as assessed by the determinants of the \lcdm\ parameter covariance matrices from fits to $2\leq\ell\leq799$ and $800\leq\ell\leq2508$, which differ by only a few per cent.

Adding the $800\leq\ell<1000$ range, including the third acoustic peak, to the high-multipole \planck\ fit has a significant effect on several parameters, including $n_s$ and $\Omega_bh^2$, tightening constraints on these parameters by factors of four and two, respectively. Conversely, the uncertainty on $\theta_{\rm MC}$ is increased by 50\% for $\ell\leq800$ compared to $\ell\leq1000$. Despite these changes, the tensions discussed above for a split at $\ell=1000$ remain for a split at $\ell=800$, with the $2.5\sigma$ tension in $\Omega_ch^2$ for the $\ell=1000$ split shifting to $2.7\sigma$ for the $\ell=800$ case (assuming a $\tau=0.07\pm0.02$ prior). From $\ell\geq800$ we find $\sigma_8\Omega_m^{0.25}=0.657\pm0.018$, which is higher than the \planck\ $\phi\phi$ constraint in equation (1) by $2.4\sigma$, the same difference as for $\ell\geq1000$. We conclude that the particular choice of $\ell=1000$ is not driving our results.

\section{Discussion}

We have found multiple similar tensions at the $>2\sigma$ level between the \planck\ 2015 high-multipole TT power spectrum and a range of other measurements. In general such tensions could be due to: (i) statistical fluctuations, (ii) an incorrect cosmological model, or (iii) systematic errors or underestimation of statistical errors in the \planck\ spectrum. A combination of these factors is also possible.

If the tensions were largely due to an unlikely statistical fluctuation, our results suggest that it is parameters from the high-multipole \planck\ TT spectrum that have scattered unusually far from the underlying values, on the basis that the low-multipole \planck\ TT, \wmap, \planck\ $\phi\phi$, BAO and distance ladder $H_0$ measurements are all in reasonable agreement with one another \citep[see also][]{bennett/etal:2014}. One might argue that the $\ell<1000$ \wmap\ and \planck\ constraints are pulled away from the true values by the multipoles at $\ell<30$. However, all parameter constraints we have quoted include cosmic variance uncertainty and thus account for this possibility (assuming Gaussian fluctuations). Furthermore, an unusual statistical fluctuation in the $\ell<1000$ values cannot explain the disagreement between the \planck\ $\ell\geq1000$ constraints and SPT, \planck\ $\phi\phi$, BAO, and the distance ladder measurements.

Cosmology beyond standard \lcdm\ cannot be ruled out as the dominant cause of tension. We do not favor this explanation because, firstly, none of the physically motivated modifications investigated by \citet{planck/13:2015} were found to be significantly preferred in fits to the full \planck\ TT spectrum, and, secondly, the most precise measurements of the CMB damping tail, from \planck\ and SPT, disagree, as discussed in Sections~3.2 and 3.3.

From 2013 to 2015 the \planck\ results were revised due to several significant systematic effects. Without more detailed reanalysis of the \planck\ 2015 data we are not in a position to comment on remaining sources of systematic error in the \planck\ high-multipole spectrum. We do note that the TT covariance matrices described in \citet{planck/13:2015} were calculated analytically assuming that sky components are Gaussian. Both foregrounds and the primary CMB have known non-Gaussian characteristics \citep[in the latter case due to lensing, see, e.g.,][]{benoit-levy/etal:2012} that would result in this approximation underestimating the true TT spectrum uncertainties, particularly at high multipoles where the foreground power becomes comparable to the CMB signal and the lensing effect is most important.

Finally, we emphasize that, irrespective of what is responsible for these tensions, care must clearly be taken when interpreting joint fits including the full range of \planck\ multipoles, particularly given \planck's high precision and ability to statistically dominate other measurements, regardless of accuracy.

\section{Conclusions}

We have discussed tensions between the \planck\ 2015 high-multipole TT spectrum ($\ell\geq1000$, roughly the scales inaccessible to \wmap) and the cosmological measurements:

\begin{enumerate}[label=(\roman*)]
\item the \planck\ 2015 TT spectrum at $\ell<1000$, which prefers a value of $\Omega_ch^2$ $2.5\sigma$ lower than the high-multipole fit,
\item the \planck\ 2015 $\phi\phi$ lensing power spectrum, which has an amplitude (parametrized by $\sigma_8\Omega_m^{0.25}$) $2.4\sigma$ lower than predicted from the $\ell\geq1000$ TT spectrum,
\item the SPT TT spectrum, covering $650\leq\ell\leq3000$, which predicts, for example, a Hubble constant $2.7\sigma$ higher than \planck\ $\ell\geq1000$,
\item the most precise measurement of the BAO scale, from the BOSS CMASS galaxy sample at effective redshift $z=0.57$, which disagrees at the $2.5\sigma$ level, and
\item the most precise local distance ladder determination of $H_0$, which is is tension at the $3.0\sigma$ level.
\end{enumerate}
These differences are quoted assuming $\tau=0.07\pm0.02$. We found that some tensions are reduced by allowing larger values of $\tau$ but note that this would introduce new tension with \planck\ polarization data. Definitive conclusions about $\tau$ will require a more detailed understanding of low-$\ell$ foreground contamination. The Cosmology Large Angular Scale Surveyor (CLASS) is expected to provide a cosmic variance limited measurement of $\tau$ \citep{class:2014,watts/etal:2015}.

Given these results and the previously reported tensions with some weak lensing and cluster abundance data, we suggest that the parameter constraints from the high-multipole \planck\ data appear anomalous due to either an unlikely statistical fluctuation, remaining systematic errors, or both. Understanding the origin of these discrepancies is important given the role \planck\ data might play in future cosmological advances.\\

We are grateful to Adam Riess for reading the manuscript and providing helpful comments. We also thank Erminia Calabrese and Karim Benabed for help with \texttt{CosmoMC} and the \planck\ likelihood code, respectively. This research was supported in part by NASA grant NNX14AF64G and by the Canadian Institute for Advanced Research (CIFAR). We acknowledge the use of the Legacy Archive for Microwave Background Data Analysis (LAMBDA). This work was based on observations obtained with \planck\ (\url{http://www.esa.int/Planck}), an ESA science mission with instruments and contributions directly funded by ESA Member States, NASA, and Canada. Part of this research project was conducted using computational resources at the Maryland Advanced Research Computing Center (MARCC).


\end{document}